\documentclass[useAMS,usenatbib]{mn2e}
\usepackage{graphicx,amssym, subfigure}
\def\oiii{{[O \sc iii]}\,}
\def\oii{{[O \sc ii]}\,}
\def\nii{{[N \sc ii]}\,}

\newcommand{\bc}{\begin{center}}
\newcommand{\ec}{\end{center}}
\newcommand{\cf}{\ifmmode C_f\else $C_f$\fi}

\title[Boxy H$\alpha$ Emission profiles in Star-Forming Galaxies]
      {Boxy H$\alpha$ Emission Profiles in Star-Forming Galaxies}

\author[Chen et al.]{
\parbox[t]{\textwidth}{\raggedright
Yan-Mei Chen$^{1}$\thanks{Email: chenym@nju.edu.cn},
Qiu-Sheng Gu$^{1}$, 
Christy A. Tremonti$^{2}$,
Yong Shi$^{1}$,
Yifei Jin$^{1}$
}\\
\vspace*{6pt}\\
$^1$School of Astronomy and Space Science, Nanjing University, Nanjing 210093, P.R. China\\ 
    Key Laboratory of Modern Astronomy and  Astrophysics (Nanjing University), Ministry of Education, Nanjing 210093, China\\
    Collaborative Innovation Center of Modern Astronomy and Space Exploration, Nanjing 210093, China\\
$^2$Department of Astronomy, University of Wisconsin-Madison, 1150 University Ave, Madison, WI 53706, USA\\}
\begin{document}



\maketitle

\label{firstpage}

\begin{abstract}
  We assemble a sample of disk star-forming galaxies from the Sloan
  Digital Sky Survey Data Release 7, studying the structure of H$\alpha$ 
  emission lines, finding a large fraction of this sample contains boxy 
  H$\alpha$ line profiles. This fraction depends on galaxy physical
  and geometric parameters in the following way: (1) it increases 
  monotonically with star formation rate per unit area ($\Sigma_{\rm SFR}$), 
  and stellar mass ($M_*$), with the trend being much stronger with 
  $M_*$, from $\sim$0\% at $M_*=10^{10}M_{\odot}$ to about 50\% at 
  $M_*=10^{11}M_\odot$; (2) the fraction is much smaller in face-on systems 
  than in edge-on systems. It increases with galaxy inclination ($i$) while 
  $i < 60\,^{\circ}$ and is roughly a constant of 25\% beyond this range; 
  (3) for the sources which can be modeled well with two velocity 
  components, blueshifted and redshifted from the systemic velocity, 
  these is a positive correlation between the velocity difference of 
  these two components and the stellar mass, with a slope similar 
  to the Tully-Fisher relation; (4) the two components are very 
  symmetric in the mean, both in velocity and in amplitude. The four 
  findings listed above can be understood as a natural result of a 
  rotating galaxy disk with a kpc-scale ring-like H$\alpha$ emission region.
  
\end{abstract}
\begin{keywords}
   galaxies: evolution -- galaxies: star formation
\end{keywords}

\section{Introduction}
Over the last decade, a huge amount of effort has been expended mapping 
the evolution of the star formation rate (SFR) density of the universe 
with time and stellar mass
\citep[e.g.,][]{brinchmann04, bauer05, feulner05, noeske07, zheng07, chen09}.
The H$\alpha$, \oii, ultraviolet-optical 
spectral energy distribution (SED) fitting and infrared 
photometry are commonly used to estimate the SFR at different redshifts.
Two major conclusions of these studies are (1) the comoving SFR density 
of the universe has declined by an order of magnitude since $z\sim2$ \citep{hopkins06}; (2) the SFR per unit stellar mass (SSFR) 
depends on both stellar mass and redshift, with massive galaxies
forming most of their stars earlier than less massive systems 
\citep{heavens04, thomas05}. 

To fully understand galaxy formation and evolution, it is important to
understand both how the population as a whole evolves and how SFR
\emph{within} individual galaxies evolves. A first step is looking at
how SFR is distributed spatially in galaxies as a function of their
physical parameters, such as stellar mass and environment. However
this is not an easy task, especially for high$-z$ galaxies given their
small angular size and faintness.  The radial distribution of
star formation in local galaxies has been analysed in individual
galaxies \citep[e.g. NGC1566,][]{comte82} and samples of tens to
hundreds of galaxies using 
narrow band imaging centered on H$\alpha$ which traces the
ionizing photons produced by massive stars.  \citet{ryder94} studied
the relative scale lengths of H$\alpha$, $V-$ and $I-$ band emission
in 34 S0-Sm galaxies, finding that the line emission had a larger
scale length than the continuum. \citet{james09} compared the
H$\alpha$ and $R$-band light profiles of 313 S0a-Im field galaxies,
finding the major central H$\alpha$ deficit (relative to the continuum) 
happens in barred galaxies,  particularly in early-type and hence high mass 
barred galaxies \citep[see also][]{hakobyan14, nair10}.
Statistical studies of larger volumes have so far been lacking 
due to the decrease in spatial resolution with distance, and the 
necessity of using multiple narrow band filters to accommodate the 
redshifting of the H$\alpha$ line.

Sloan Digital Sky Survey provides the largest galaxy sample of photometry
and spectra so far. It is a great database to study all kinds of properties
of local galaxies. \citet{ge12} select double-peaked narrow emission-line galaxy sample
from SDSS DR7 and study their properties.
However, due to the 3" diameter fiber, it is hard to
derive any spatial information about the galaxies directly from the spectra.
In this paper, we make the first attempt of getting spatial information 
from the structures of H$\alpha$ emission lines in SDSS spectra. As we know, 
the H$\alpha$ profile depends on the strength of emission at a given velocity,
which is determined by the rotating curve and the H$\alpha$ surface brightness 
distribution in a rotating disk. Thus through comparing observed H$\alpha$ line 
structures with the simulated H$\alpha$ line profiles for the rotating disk model, 
we can have an idea of the possible distribution of the H$\alpha$ surface 
brightness. In \S2, we introduce the 
sample selection criteria used in our study. We characterize the H$\alpha$
emission line profiles and study the relation between line profiles and  
galaxy physical/geometric parameters in \S3. The origin of the 
boxy line profile (please see \S3.1 for the definition of peakiness or boxiness 
of a line) is found to be a rotating disk with a kpc-scale ring-like H$\alpha$
emission region in \S4. The results are summarized in \S5. 

\section{Sample and Data Analysis}
\subsection{The Data \label{data}}

The seventh data release \citep[DR7;][]{abazajian09} of Sloan Digital
Sky Survey \citep[SDSS;][]{york00} contains $\sim$930,000 galaxy
spectra. The spectra are taken through $3^{\prime\prime}$ diameter
fibers with a dispersion of 69~km~s$^{-1}$~pixel$^{-1}$. These
galaxies cover a redshift range of $z=0-0.3$.
In this work, we analyze objects drawn from the ``Main" galaxy sample
\citep{strauss02} which are selected to have Petrosian magnitudes
in the range $14.5 < r< 17.7$ after correction for foreground Galactic
extinction using the reddening maps of \citet{schlegel98}.

The stellar continuum is fitted with stellar population models
\citep{tremonti04, brinchmann04}. The basic assumption of this fitting
is that any galaxy star formation history can be approximated as a sum
of discrete bursts. The library of template spectra is composed of
single stellar population (SSP) models generated using a preliminary
version of the population synthesis code of Charlot \& Bruzual (in
prep., hereafter CB08). We refer the reader to \citet{tremonti04} for
more details about the fitting process. The results of this fitting
procedure and measurements of a number of line indices (e.g., D4000, 
H$\delta_A$) can be found in the MPA/JHU catalogue\footnote{The MPA/JHU 
  catalog can be downloaded from
  http://www.mpa-garching.mpg.de/SDSS/DR7.}. 

We use our best fit stellar continuum model to re-measure each
galaxy's redshift using a cross correlation technique and masking out
regions of the data with emission lines. These redshifts differ very
slightly from the SDSS pipeline redshifts which use cross correlation
templates that include emission lines. The new redshifts enable us to
accurately measure the gas motions with respect to the stars.

The derived galaxy parameters required in this work include stellar
mass ($M_*$), star formation surface density
($\Sigma_{\rm SFR}$), and galaxy inclination ($i$). The stellar masses
are estimated from the $ugriz$ colors of the galaxies. Connecting these
colors with a large grid of CB08 model colors following the
methodology described in \citet{salim07}, the maximum likelihood
estimate of the $z$-band mass-to-light ratio for a galaxy can be
obtained. We derive SFRs from the dust extinction corrected H$\alpha$
luminosity using the formula of \citet{kennicutt98}, converting 
from a Salpeter to Kroupa initial mass function \citep{kroupa01} by
dividing by a factor of 1.5. $\Sigma_{\rm SFR}$ is defined as SFR/$\pi
R^2$, where $R$ is the $1.5^{\prime\prime}$ fiber radius of SDSS
corrected for projection effects. We use  $\Sigma_{\rm SFR}$
instead of SFR because the former is less subject to aperture bias.

The SDSS photometric pipeline \citep{lupton01} fits a two-dimensional
model of an exponential profile and a \citet{devau48} profile
to each galaxy image. The best linear combination of the exponential
and de Vaucouleurs models is stored in a parameter called $\rm
fracDeV$ \citep{abazajian04}. An axial ratio ($b/a$) is given by this
fitting procedure. Here we use $\rm fracDeV$ to distinguish disk
galaxies (${\rm fracDeV}<0.8$) from early-type galaxies (${\rm
  fracDeV}\ge0.8$) \citep{padilla08}. The axis ratios from the
exponential and de Vaucouleurs models are consistent with each other.
The inclinations of disk galaxies are computed from the measured axial
ratio, $b/a$, and the $r-$band absolute magnitude $M_{\rm r}$ using
Table 8 in \citet{padilla08}.

\subsection{Sample Selection \label{sample_selection}}
The sample is selected by the following criteria:

\begin{enumerate}
\item redshift of $\rm 0.1 \le z \le 0.15$. We limit our sample to
  this small redshift range to ensure that we are sampling the same
  physical scale of the galaxies. With $3^{\prime\prime}$ diameter fibers, the
  SDSS spectra probe the central 5--7 kpc in this redshift range.

\item $r-$band $\rm fracDeV < 0.8$. This criterion selects 
disk galaxies, which makes the calculation of the inclination angle 
from $b/a$ possible. 

\item $\log (\Sigma_{\rm SFR}) > -0.15$. This criterion selects
  galaxies that are forming stars. Combined with our redshift cut,
  it also insures that the H$\alpha$ lines are measured with reasonable 
  signal-to-noise (S/N). 

\item $\log ($\oiii/H$\beta) <  0.61/\{\log ($\nii/H$\alpha)-0.05\} +1.3$.
Only star forming galaxies are included in our sample \citep{kauffmann03}. 
\end{enumerate}

We refer to this sample hereafter as our {\em parent} sample.  It contains
18,425 galaxies. 

\begin{figure*}
\bc
\hspace{-0.1cm}
\resizebox{17cm}{!}{\includegraphics{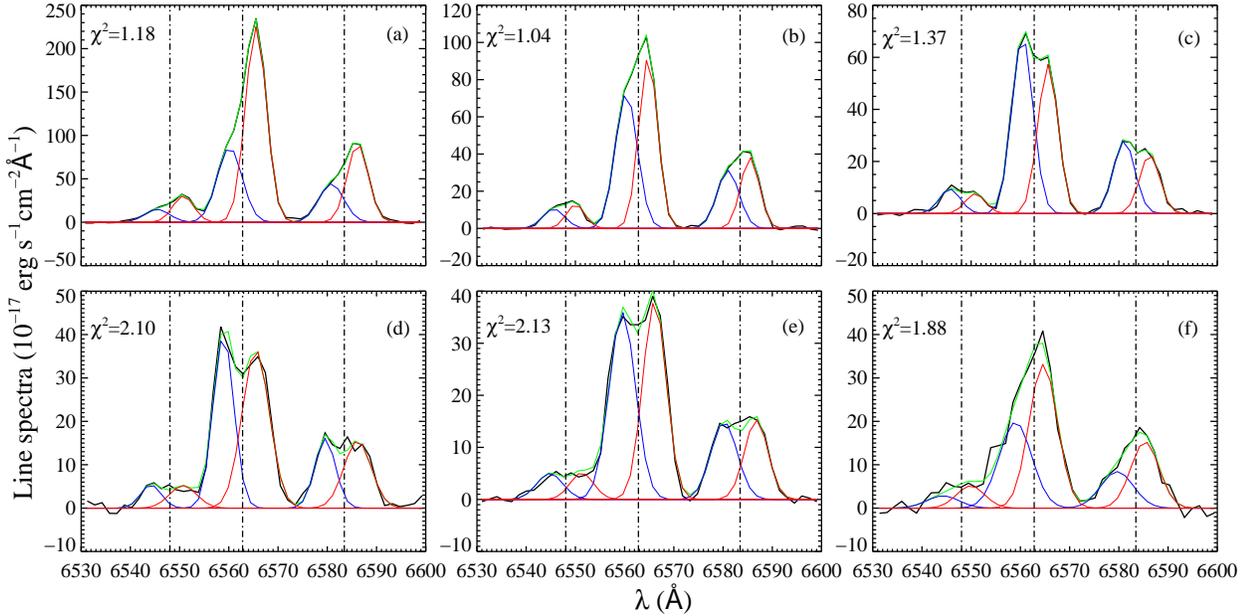}}\\%
\caption{Examples of boxy H$\alpha$ and \nii emission. The data is shown in black;
blue and red lines show Gaussian fits to the blueshifted and 
redshifted emission components; the green lines show the combined fits.
The dashed vertical lines mark the restframe wavelength of the \nii $\lambda$6548,
H$\alpha$ $\lambda$6563, \nii $\lambda$6583 emission lines. The top panel shows cases 
that can be modeled well with two Gaussian components; the bottom panel shows failed fits 
due to complicated line structures.  The reduced $\chi^2$
is given in the top-left corner of each panel. \label{Fig_spec_examp}}
\ec
\end{figure*}

\section{H$\alpha$ Emission Line Profiles \label{boxy_frac}}
\subsection{Characterizing Emission Line Profiles}
We calculate the fourth moment or {\em kurtosis} of the H$\alpha$ emission
lines (hereafter ${\rm H}\alpha_{\rm kur}$) from the continuum-subtracted
spectrum. The kurtosis characterizes the peakiness or boxiness of a
line. For a gaussian profile, we would expect a ${\rm H}\alpha_{\rm
  kur}=3$, with boxier profiles yielding smaller values of 
kurtosis. We find that sources with ${\rm H}\alpha_{\rm kur} < 2.4$
contain more than one velocity component. 

We also explore parametric fits to the line profiles. We use six gaussian
components to fit the H$\alpha$ and \nii $\lambda$6548,6584 lines
simultaneously, with two components for each line. Each \nii
$\lambda$6548 component is forced to have the same centroid and width
as the corresponding \nii $\lambda$6583 component, and the flux ratio
of \nii $\lambda$6583 to \nii $\lambda$6548 is fixed to be 3. The top
panel of Figure~\ref{Fig_spec_examp} shows three typical examples of
our fits. In some cases, the H$\alpha$ lines can be fitted well with
two gaussian components, one blueshifted and the other redshifted
relative to the systemic velocity defined by the stars. There are
other cases in which two gaussian components cannot model the line
profile due to the complicated structure of H$\alpha$ (see the buttom panel). 
The reduced $\chi^2$ is given in the top-left corner of each panel.

For comparison, we visually inspect all the emission line spectra and try
to select sources with boxy H$\alpha$ emission by eye. In this sample
we only include objects which clearly contain more than one velocity
component in both the H$\alpha$ and \nii lines and we require the
components to be blueshifted and redshifted relative to the systemic
velocity. We do not include galaxies that exhibit low-level wings on
their line profiles.  As we will show in \S\ref{boxy_frac}, we find
good agreement between our visually identifed sample and galaxies with
${\rm H}\alpha_{\rm kur} < 2.4$. We refer to the 2815 sources
with ${\rm H}\alpha_{\rm kur} < 2.4$ as ``{\em boxy sources}" hereafter.

\begin{figure*}
\bc
\hspace{-0.1cm}
\resizebox{17cm}{!}{\includegraphics{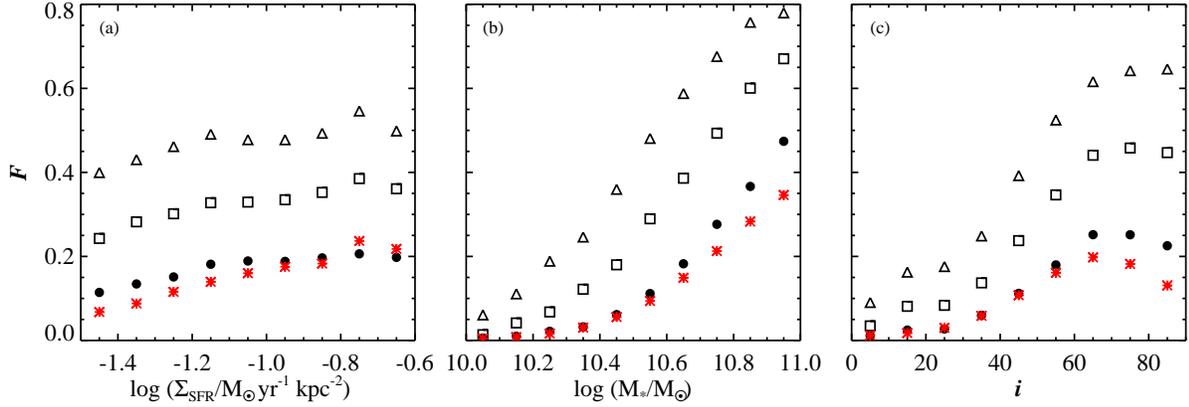}}\\%
\caption{The fraction of boxy sources vs. star formation surface
  density, $\Sigma_{\rm SFR}$, stellar mass, $M_{*}$, and galaxy 
  inclination, $i$. The black points indicate the fraction of galaxies 
  with H$\alpha$ kurtosis below a certain threshold. For reference, a 
  Gaussian line profile has ${\rm H}\alpha_{\rm kur}=3$; smaller
  values indicate increasing boxiness. The
  triangles, squares and solid dots show 
$F=N({\rm H}\alpha_{\rm kur} < 2.7)/N_{\rm bin}$, 
$F=N({\rm H}\alpha_{\rm kur} < 2.55)/N_{\rm bin}$ and 
$F=N({\rm H}\alpha_{\rm kur} < 2.4)/N_{\rm bin}$, respectively. 
$N({\rm H}\alpha_{\rm kur} < 2.4)$ is the number of galaxies with 
${\rm H}\alpha_{\rm kur}$ smaller than $2.4$ in a given 
galaxy parameter bin and $N_{\rm bin}$ is the total number of galaxies
in the bin. Red asterisks show the fraction of galaxies visually identified as
having two velocity components.  
\label{Fig_frac_ha}}
\ec
\end{figure*}

\subsection{The Fraction of Galaxies with Boxy 
H$\alpha$ Emission Profiles\label{boxy_frac}}

In this section, we examine how the boxy source fraction varies as a 
function of galaxy physical and geometrical properties: 
$\Sigma_{\rm SFR}$, $M_*$, and $i$.

We divide our parent sample into bins in $\Sigma_{\rm SFR}$, $M_*$,
and $i$. In each bin we measure the fraction, $F$, of galaxies with
H$\alpha$ kurtosis below a given theshold, for example, $F=N({\rm
  H}\alpha_{\rm kur} < 2.4)/N_{\rm bin}$. Here $N({\rm H}\alpha_{\rm
  kur} < 2.4)$ is the number of galaxies with ${\rm H}\alpha_{\rm
  kur}$ smaller than $2.4$ in a given bin and $N_{\rm bin}$ is the
total number of galaxies in the bin. In Figure~\ref{Fig_frac_ha} we
show how the fraction of galaxies with boxy H$\alpha$ profiles varies
with galaxy physical parameters. The black symbols show the fraction
of galaxies in each bin with ${\rm H}\alpha_{\rm kur}$ less than 2.6,
2.55, and 2.4 (triangles, squares, and circles) and the red asterisks
show the fraction of galaxies in each bin visually identified as
having two velocity components. The trends between $F$ and galaxy
parameters are very similar for the different ${\rm H}\alpha_{\rm
  kur}$ theresholds, with the main differenence being in the absolute
value of $F$. The results from our visual inspection agree closely
with galaxies with ${\rm H}\alpha_{\rm kur} < 2.4$ (solid circles).
This suggests that our kurtosis measurements are very effective at
selecting galaxies with boxy H$\alpha$ profiles and that our results
are independent of the exact ${\rm H}\alpha_{\rm kur}$ threshold
adopted.

Figure~\ref{Fig_frac_ha}$a$ and $b$ show that the fraction of
galaxies with boxy H$\alpha$ lines increases monotonically with both
$\Sigma_{\rm SFR}$ and $M_*$. The fraction increases more dramatically
with stellar mass, from $\sim$0\% at $M_*=10^{10}M_{\odot}$ to about 50\% at
$M_*=10^{11}M_\odot$ for the subsample with the highest degree of boxiness
(${\rm H}\alpha_{\rm kur} < 2.4$). Figure~2c shows that the boxy fraction
increases rapidly with inclination while $i<60\,^{\circ}$. Above this
value it is roughly constant at $F\sim$25\% for the galaxies with the
highest degree of boxiness. 

\begin{figure}
\bc
\hspace{-0.1cm}
\resizebox{8.5cm}{!}{\includegraphics{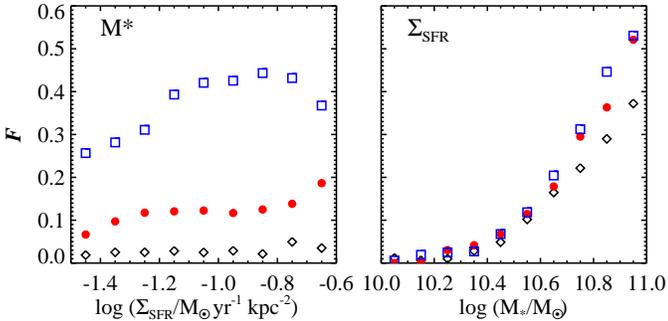}}\\%
\caption{The dependence of the boxy source fraction on
  $\Sigma_{\rm SFR}$ and $M_*$. In each panel, the sample is divided
  into three sub-bins by the parameter marked in the top-left corner.
  The black, red, and blue points indicate the low, median, and high
  sub-bins respectively. This approach is designed to help mitigate
  the effect of $M_*-\Sigma_{\rm SFR}$ correlation. The physical parameters
  that play the dominant role in driving the trends will show the
  smallest offsets between the black, red, and blue points.
\label{Fig_frac_bin}}
\ec
\end{figure}

Figure ~\ref{Fig_frac_ha} shows that the boxy source fraction depends
more strongly on $M_*$ than on $\Sigma_{\rm SFR}$. However, due to the
correlation between $\Sigma_{\rm SFR}$ and $M_*$ for star forming disk
galaxies (see Figure 1 of Chen et al. 2010), it is difficult to figure
out whether $\Sigma_{\rm SFR}$ is a real driver of boxy line profiles
or whether the trend between $F$ and $\Sigma_{\rm SFR}$ is just an
inevitable by-product of the $F-M_*$ relation. To solve this problem,
we split $\Sigma_{\rm SFR}$ into different bins and take each bin in
$\Sigma_{\rm SFR}$ and futher divide it into three equal sub-bins,
sorting the galaxies by $M_*$. We are then able to study the trends
between $F$ and $\Sigma_{\rm SFR}$ in low, median and high $M_*$ bins
(where the exact division between low, median and high changes with
$\Sigma_{\rm SFR}$). If $\Sigma_{\rm SFR}$ is the most important
parameter in driving the boxy line profiles, then very little
difference is expected between the sub-bins in $M_*$ and fixed
$\Sigma_{\rm SFR}$. Conversely, if $M_*$ is the dominant driver, the
three sub-bins will be strongly offset from one another at each value
of $\Sigma_{\rm SFR}$.

Figure~\ref{Fig_frac_bin} shows the fraction of boxy sources
($F=N({\rm H}\alpha_{\rm kur} < 2.4)/N_{\rm bin}$) as a function of $\Sigma_{\rm
  SFR}$ and $M_*$. In each panel the sample is split into three
sub-bins according to the galaxy parameter labeled in the top-left
corner. The black, red, and blue points indicate the low, median and
high sub-bins respectively. Figure~\ref{Fig_frac_bin}
adds to the evidence that $M_*$ is the dominant driver of the
boxy line profiles and proves that the $F-\Sigma_{\rm
  SFR}$ relation shown in Figure~\ref{Fig_frac_ha} is a by-product of
$F-M_*$ and $M_*-\Sigma_{\rm SFR}$ correlation. The $F-M_*$ trend
appears to be totally independent of $\Sigma_{\rm SFR}$ at low masses,
with a very weak trend evident at $M_* > 10^{10.7} M_\odot$.

\section{The Origin of  Boxy H$\alpha$ Profiles}

Having established that boxy H$\alpha$ profiles are common in massive
edge on disk galaxies in the SDSS, we now turn to the question of their
origin.

\subsection{Bi-polar Outflows}

H$\alpha$ emission line profiles have been studied in detail in
several nearby starburst galaxies using longslit spectra or
narrow band images \citep[e.g.,][]{heckman90, lehnert96, greve00,
  west09}. The purpose of these works was to search for evidence of
stellar wind/supernova driven outflows and to constrain their geometry
and dynamics. In these works, double-peaked H$\alpha$
emission-line profiles (one special case of boxy structure) were
commonly found in minor axis spectra. This was explained as a
combination of the emission from both the near side and far side of
one outflow cone.

An SDSS fiber spectrum samples the central $3^{\prime\prime}$ of a
galaxy which corresponds to radii of 2.5 -- 3.5~kpc for our sample. 
This aperature could encompass the emission from both the disk and 
two outflow bi-cones if they exist. However, it should be kept in 
mind that the emission from the outflow is expected to be very weak 
relative to the disk. \citet{lehnert99} looked at H$\alpha$ emission 
in the starburst galaxy M82 and found that the wind contributes only 
$\sim0.3$\% of the total flux.  The other arguement against the
outflow hypothesis is that the fraction of boxy sources correlates much
more strongly with $M_*$ than $\Sigma_{\rm SFR}$.  This is surprising
because \citet{chen10} showed that the amount of cool gas in galactic 
winds in normal star forming galaxies is a strong function 
of $\Sigma_{\rm SFR}$.  Thus we conclude
that bi-polar galactic winds, while possibly present, are not
responsible for the boxy H$\alpha$ line profiles we observe. 

\begin{figure}
\bc
\hspace{-0.1cm}
\resizebox{8.5cm}{!}{\includegraphics{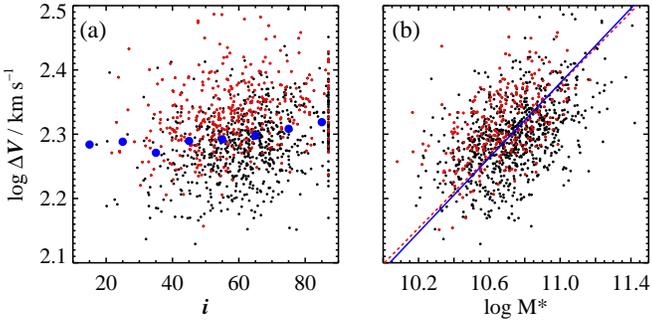}}\\%
\caption{The dependence of $\Delta v$, the velocity difference between the two 
components, on inclination and stellar mass. The black dots show the
data points. In the left panel, the blue dots are medians. In the right
panel, the blue line is a linear regression of the
data. The red dashed line has the same slope as the Tully-Fisher
relation and an arbitrary intercept. The over-plotted red dots represent 
high mass ($M_*>10^{10.8}M_\odot$) and high inclination ($i>70\,^{\circ}$) 
sources in left and right panels respectively. 
\label{split_inc_mass}}
\ec
\end{figure}
\begin{figure}
\bc
\hspace{-0.1cm}
\resizebox{8.5cm}{!}{\includegraphics{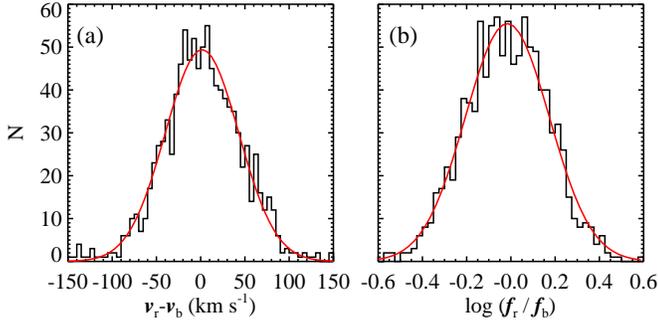}}\\%
\caption{Histograms of the velocity difference ($v_{\rm r}-v_{\rm b}$;
  left panel) and flux ratios ($f_{\rm r}/f_{\rm b}$; right
  panel) between the blue-shifted and red-shifted emisison line components.
  The red curves show gaussian fits.\label{hist_v_flux}}
\ec
\end{figure}

\subsection{Star Formation in the Disk}
The boxy H$\alpha$ line profiles may also be produced by star
formation in the disk for certain H$\alpha$ surface brightness 
distributions. The simplest case is star forming disk with 
a central hole.  In this scenario,
the blueshifted and redshifted velocity peaks arise from gas in the
ring moving directly towards and away from us. While the bi-polar
outflow model encounters problems in explaining the trends between the
boxy source fraction and galaxy physical parameters ($\Sigma_{\rm
  SFR}$ and $M_*$), these two relations can be naturally explained by
the rotating disk model: the higher the stellar mass, the larger the
rotational velocity of the disk, which would lead to larger velocity
splitting and a higher detection rates of boxy sources.

The inclination dependence is also broadly consistent with our
expectations for a disk.  The velocity spread between the peaks 
will scale as sin($i$) and be maximum when the disk is edge on
($i=90^{\circ}$).  At lower inclinations only the 
most massive galaxies with the largest intrinsic rotation speeds
will be detected as boxy sources, hence the fall off in $F$ at low
$i$. Above $i=60\,^{\circ}$ inclination effects change the velocity
spread by less than 15\%, so $F$ appears roughly constant. The
small dip at $i\sim90$ may be due to extinction effects in 
fully edge-on disks.

For the 1088 sources that can be modeled well with two velocity
components, we examine how the velocity difference between the peak of
the two gaussians ($\Delta v$) depends on galaxy inclination and
stellar mass. Figure~\ref{split_inc_mass}$a$ shows $\Delta v$ as a
function of $i$, with blue dots indicating median values.  At first
sight, the lack of correlation in this plot conflicts with the
prediction of the toy rotating disk model: $\Delta v \propto$
sin($i$). However, there is a strong selection effect at work: it is
not possible to fit two well-separated gaussians to sources with
$\Delta v$ below the SDSS resolution of 150~km~s$^{-1}$.  At fixed
inclination we expect a range in mass and thus a range in $\Delta v$
down to the resolution limit.  
The over-plotted red dots in Figure~\ref{split_inc_mass}$a$
show the subsample of high mass galaxies with $M_*>10^{10.8}M_\odot$,
which is believed to suffer less selection effect.

Figure~\ref{split_inc_mass}$b$ shows that  
$\Delta v$  increases with increasing $M_*$. 
The blue line shows a linear regression:
\begin{equation}
\log\Delta v = (0.29\pm 0.02)\log{M_*} - (0.81\pm 0.18)
\end{equation}
The fit has a Pearson's correlation coefficient of 0.53 and a
probability of less than $10^{-5}$ that it could be obtained by
chance. The red dashed line has the same slope as the Tully-Fisher
relation \citep{tully77, bell01} with an arbitrary value of the intercept.  
Since most of the objects in Figure ~\ref{split_inc_mass}$a$ have 
$i>50\,^{\circ}$, the effects of inclination on $\Delta v$ are 
small enough to be neglected in Figure~\ref{split_inc_mass}$b$. The
similarity of the slope of our $\Delta v$--$M_*$ relation and the
Tully-Fisher relation is somewhat surprising considering that the SDSS 
fiber apertures do not necessarily sample out to the flat part of the
rotation curve. Nevertheless, this consistency provides support
for the rotating disk model as the dominant origin of the boxy line
profiles found in our sample. We also over-plotted the high inclination 
sources ($i>70\,^{\circ}$) in Figure~\ref{split_inc_mass}$b$ as red dots 
to show how the selection effect of inclination influences our result. 

\subsubsection{Further Tests of the Disk Hypothesis}

The rotating disk model also predicts that the blueshifted and
redshifted components should have roughly equal and opposite
velocities relative to the systemic velocity of the stars ($v_r - v_b
= 0$ km~s$^{-1}$). Moreover, while patchy extinction could result in
one velocity component being stronger than the other in an individual
galaxy, in the mean, the flux ratio of the two components should be
unity ($f_b/f_r=1$). We test these two predictions in Figure~\ref{hist_v_flux}.
Figure~\ref{hist_v_flux}$a$ shows a histogram of the velocity difference 
between the redshifted ($v_{\rm r}$) and blueshifted ($v_{\rm
  b}$) components.  As expected, for the disk model, the distribution
is strongly peaked about zero. Figure~\ref{hist_v_flux}$b$ shows the 
distribution of flux ratios between the redshifted ($f_{\rm r}$) and 
blueshifted ($f_{\rm b}$) components. It is also very consistent with 
the expectations of the disk model. 

\subsection{Evidence for central kiloparsec-scale H$\alpha$ deficit region}

\subsubsection{Simple Models}
\begin{figure}
\bc
\hspace{-0.1cm}
\resizebox{8.5cm}{!}{\includegraphics{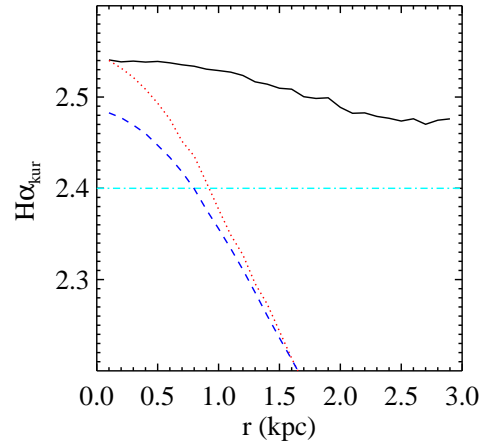}}\\%
\caption{Kurtosis vs. the hole size at fixed circular velocity under a certain 
assumption of H$\alpha$ light profile distribution. The black-solid curve 
shows an exponential profile which flattens in the center, $r$ is the radius of 
the central flat area. Red-dotted and blue-dashed curves represent an exponential and a flat 
profile with an inner hole, respectively. In these two cases, $r$ is the radius 
of the hole. The cyan horizontal line marks our boxy source selection criterion 
${\rm H}\alpha_{\rm kur} = 2.4$. \label{kur_r}}
\ec
\end{figure}

Having demonstrated that the boxy H$\alpha$ emission line profiles
arise from star formation in a rotating disk, we now turn to
discussing the distribution of star formation.  To do this we
  employ a simple toy model of a fully edge-on disk in solid body
  rotation out to 3 kpc and a flat rotation curve with 
  $V_C$=200~km~$s^{-1}$ at larger radii. 
 The observed H$\alpha$ line profile depends on the strength of the
emission at a given velocity.  To determine this we consider various
H$\alpha$ surface brightness distributions and add up the amount of
emission at each velocity within 3 kpc (the SDSS aperture). The 
resulting H$\alpha$ line profile is then
convolved to the SDSS spectral resolution of 150 km~s$^{-1}$. 
  For an exponential H$\alpha$ light distribution, the synthetic
  profile has ${\rm H}\alpha_{\rm kur} = 2.55$, which is less peaked
  than a Gaussian (${\rm H}\alpha_{\rm kur}=3$) but considerably more
  peaked than line profiles meeting our boxy source criterion 
(${\rm H}\alpha_{\rm kur}=2.4$).  We experiment with three other cases of
the light distribution: (1) an exponential profile with a hole (no
H$\alpha$ emission) in the center. The red-dotted curve in Figure~\ref{kur_r}
shows how kurtosis maps to the size $r$ of the hole in our toy model;
(2) a flat H$\alpha$ profile with an inner hole which has a radius of
$r$ (blue-dashed curve in Figure~\ref{kur_r}); (3) a profile that is
exponential and then flattens inside radius $r$ (black-solid curve in
Figure~\ref{kur_r}. The horizontal cyan line marks our boxy source selection
criterion ${\rm H}\alpha_{\rm kur} = 2.4$. To get a boxy H$\alpha$
profile a central H$\alpha$ emission deficiency appears to be
necessary. 
Both the blue-dashed and red-dotted curves in Figure~\ref{kur_r} suggests
that to satisfy the boxy source criterion, the radius of the deficit region
should be $\sim$1 kpc. We find that the exact size of the deficit region 
required for the boxy profile depends on the steepness of the rotation curve and the size
of the observation aperture, steeper rotation curve or larger 
aperture leads to stronger boxiness. After exploring a range of reasonable 
parameters for these variables we conclude that a kpc-scale central 
H$\alpha$ deficit region is necessary to produce the boxy H$\alpha$ structure.

In the model discussed above, we did not sample the flat part of the rotation curve.
We keep in mind that we can get a boxy H$\alpha$ profile when fiber aperture samples 
out to the flat part of the rotation curve since a lot of H$\alpha$ comes out at a single velocity.
Figure 1 of  \citet{catinella06} shows that the radius where the turn over of rotation curve happens
is around $r_d$ for the two brightest bins, and for the other less luminous bins, this radius is larger than
$r_d$, where $r_d$ is the exponential disk scale length. However, more than 80\% galaxies with boxy 
H$\alpha$ profiles studied in this work have $r_d$ larger than the fiber radius, indicating that we are 
not going to see very much gas that is on the flat part of the rotation curve. On the other hand, we further check 
the fraction of galaxies with boxy H$\alpha$ profiles in bins of stellar mass with each mass bin split into three 
bins of redshift. There is no evidence that the more distant galaxies have higher frequency of boxy H$\alpha$ 
profiles, suggesting that the larger physical size of the fiber was not playing a role in the frequency of boxy H$\alpha$
profiles. In summary, targeting the flatten part of the rotation curve could not be the dominate reason for the boxy H$\alpha$ 
profiles.

\subsubsection{Dust Obscuration}

An interesting question is whether these H$\alpha$ deficit regions are due to a
central deficit of star formation or whether they could be due to the
effects of dust obscuration.  Based on a sample of 10,095 galaxies
with bulge$-$disc decompositions in the Millennium Galaxy Catalogue,
\citet{driver07} infer a large amount of dust in the inner parts of
galaxies.  They conclude that 71\% of all $B-$band photons produced in
bulges in the nearby Universe are absorbed by dust.  To test the
extinction hypothesis for our H$\alpha$ holes, we selected galaxies
with boxy H$\alpha$ profiles (${\rm H}\alpha_{\rm kur} < 2.4$) and
high inclination and stellar mass ($i>60\,^{\circ}$, log
$M_*/M_\odot>$10.8). The projected disk rotation velocities of
the galaxies in this sample are larger than the SDSS instrumental
resolution, enabling us to measure dust attenuation as a
function of velocity.  The spectra of the galaxies were normalized to the
median flux between 5450 and 5500~\AA\, where the spectrum is free of
strong absorption and emission lines, averaged in the restframe to
increase the S/N, and fit with a stellar population synthesis model
(see \S2.1 for more details).  We then examined the
continuum-subtracted H$\alpha$ and H$\beta$ line profiles as a function of
velocity. Dust attenuation, traced by the H$\alpha$$/$H$\beta$
ratio, is observed to vary with velocity, but only weakly. The mean
attenuation correction changes by only 0.1~mag from the disk center
($v=0$ km~s$^{-1}$) to the outer regions ($v > 150$~km~s$^{-1}$).
While our limited velocity resolution is clearly an issue, this change 
is still much smaller than the amount of attenuation needed to cause the
central H$\alpha$ deficit.

\section{Summary}

We study the structure of H$\alpha$ emission lines in a disk star
forming galaxy sample drawn from the SDSS DR7, deriving  
information on star formation from the H$\alpha$ emission lines by comparing the 
observed line structures and the simulated line profiles.
A large fraction of boxy sources, which is identified by their kurtosis 
using the criterion ${\rm H}\alpha_{\rm kur} < 2.4$, is found from this 
sample, and this fraction increases with galaxy inclination and flattens 
at $i=60\,^{\circ}$. Although this trend can be understood in both bi-polar 
outflow and rotating disk models, three lines of evidence strongly support a disk
origin: (1) the boxy source fraction depends more strongly on stellar
mass than $\Sigma_{\rm SFR}$; (2) the velocity difference between the
two emission components scales strongly with stellar mass with a slope
very similar to the classic Tully-Fisher relation; (3) on average the
line profiles are very symmetric. 

In the rotating disk scenario, a ring-like H$\alpha$ surface
brightness distribution, namely, a kpc-scale central H$\alpha$ emission
deficient area, is required to produce the boxy line profile. The
  high fraction $\sim 50$\% of boxy sources 
  in high mass galaxies indicates that
  the H$\alpha$ hole is a common feature. We can not comment on whether 
  this behaviour extends to lower luminosity galaxies because we are limited by the SDSS
  spectral resolution. The MaNGA survey (Mapping Nearby Galaxies at Apache 
  Point Observatory, \citet{bundy15}) plans to obtain integral-field spectroscopy 
  of a representative sample of about 10,000 galaxies above stellar masses
of $10^9$$M_\odot$ with redshift around 0.03. Considering the fairly large fraction 
  of galaxies with boxy H$\alpha$ profiles in massive galaxies, we should expect to 
  see the H$\alpha$ holes in a substantial fraction of MaNGA galaxies. And we will 
  have more information to start investigating the formation mechanism of the H$\alpha$ holes, 
  e.g., star formation suppression associated, with bars, bulges, radio AGN feedback, etc.  
    
\section*{acknowledgements}
We are very grateful to referee Philip James for useful comments and 
suggestions  that  have  strengthened  this  work.  We  also  thank  Cheng Li 
for helpful discussions. Y.M.C acknowledges support from NSFC grant 11573013, the Opening Project of Key 
Laboratory of Computational Astrophysics, National Astronomical Observatories, 
Chinese Academy of Sciences. Q.S.G acknowledges support from NSFC grants
11363001, 11273015 and 11133001, National Basic Research Program (973 program No. 2013CB834905.  
C.A.T. acknowledges support from National Science Foundation of the United States Grant No. 0907839.
Y.S. acknowledges support from  NSFC  grant  11373021,  the CAS Pilot-b grant No. XDB09000000
and Jiangsu Scientific Committee grant BK20150014.

Funding for the SDSS and SDSS-II has been provided by the Alfred P. Sloan 
Foundation, the Participating Institutions, the National Science Foundation, 
the U.S. Department of Energy, the National Aeronautics and Space 
Administration, the Japanese Monbukagakusho, the Max Planck Society, and 
the Higher Education Funding Council for England. 
The SDSS Web Site is http://www.sdss.org/.

The SDSS is managed by the Astrophysical Research Consortium for 
the Participating Institutions. The Participating Institutions are 
the American Museum of Natural History, Astrophysical Institute Potsdam, 
University of Basel, University of Cambridge, Case Western Reserve 
University, University of Chicago, Drexel University, Fermilab, the 
Institute for Advanced Study, the Japan Participation Group, Johns Hopkins 
University, the Joint Institute for Nuclear Astrophysics, the Kavli 
Institute for Particle Astrophysics and Cosmology, the Korean Scientist 
Group, the Chinese Academy of Sciences (LAMOST), Los Alamos National 
Laboratory, the Max-Planck-Institute for Astronomy (MPIA), the 
Max-Planck-Institute for Astrophysics (MPA), New Mexico State University, 
Ohio State University, University of Pittsburgh, University of Portsmouth, 
Princeton University, the United States Naval Observatory, and the 
University of Washington.

\bibliographystyle{mn2e}
\bibliography{ms}

\end{document}